\documentclass[aps,prl,twocolumn,groupedaddress]{revtex4}

\usepackage{graphicx}
\begin{document}


\title{The Feynman Propagator from a Single Path}


\author{G. N. Ord}
\email[ corresponding author  ]{gord@acs.ryerson.ca}


\affiliation{M.P.C.S. \\ Ryerson University\\Toronto Ont.}
\author{J. A. Gualtieri}
\email[]{gualt@peep.gsfc.nasa.gov}
\affiliation{Applied Information Sciences Branch\\ Global Science and
Technology\\Code 935 NASA/Goddard Space Flight Center\\Greenbelt MD.}

\date{September 19 2001}

\begin{abstract}
 We show that it is possible to construct the
Feynman Propagator for a free particle in one dimension, without
quantization, from a single continuous space-time path.
\end{abstract}


\maketitle

The Feynman
path-integral formulation of Quantum Mechanics\cite{Feynman48a,FeyHibbs}
is well known  for its utility and  intuitive appeal.  An interesting
history of its development
 may be found in the article and book by
Schweber\cite{Schweber86,Schweber94}. Although the mathematics of the path
integral encourages us to think of the paths in terms of real space-time
trajectories, and there have been very interesting proposals for testing
the reality of the paths \cite{kroger97a,kroger97b,kroger00}, the
formulation itself falls short of providing a full microscopic basis for
quantum mechanics. This is in contrast to the Wiener integral
 which is an abstraction of the  microscopic model (Brownian motion) 
supporting the diffusion equation. In particular, Wiener paths are known
 to approximate actual physical trajectories of diffusing particles,
whereas the relation between Feynman paths and physical particles is not
so direct.

There are two main barriers to an association between Feynman paths and
any physical trajectory of a real particle.   First of all there is
a many-to-one correspondence between Feynman paths and the particle
being described. Interference effects require this non-uniqueness since
individual trajectories carry variable phase but not variable amplitude
in the propagator \cite{krogernote}. Thus a physical
particle cannot simply traverse a single Feynman path while propagating in
space-time.

A second impediment is that, in the path integral formulation, the
required reduction of wave functions on measurement is grafted onto the
dynamics of propagation; it does not follow in a direct fashion from the
paths themselves. As in other formulations of quantum mechanics we need
measurement postulates to interpret the theory in terms of the real
world. 
 
In this paper we  show that in the particular case of the Feynman
Chessboard model, one can modify the formulation so that the propagator
can be constructed by a single continuous space-time curve. This is done
by allowing particles to have  trajectories with reversed time segments.
Although this might seem conceptually `expensive', allowing this feature
explicitly provides the physical mechanism which creates the phase of a
wave function {\em without invoking  an analytic continuation}.  The propagator
appears naturally as a pattern created by the (space-time) plane-filling
path of a single point-particle. In the new formulation, the many-to-one
aspect of Feynman paths is circumvented  by sewing together an ensemble
of Chessboard paths into a single curve in such a way that formal
quantization is unnecessary.

  The Chessboard or Checkerboard
model\cite{FeyHibbs,Gersch81,JacSchulman84} extended Feynman's path
integral approach to the relativistic domain in order to incorporate
electron spin. In this model, particles hop with speed $\pm c$ on a
discrete space-time lattice with spacing
$\epsilon$. Choosing units in which $c=1$,  paths consist of diagonal
segments resembling forward bishop's moves in chess(Fig. 1). 

\begin{figure}
\includegraphics[scale = .42]{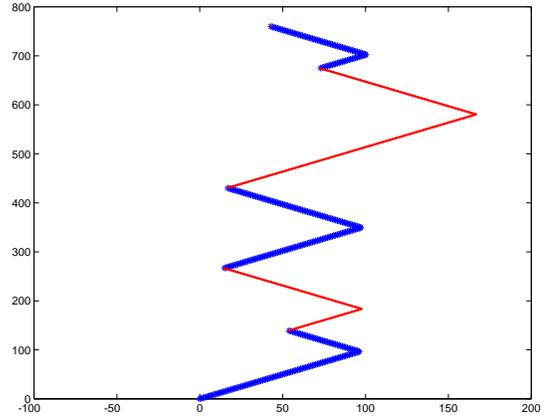}
\caption{A Feynman Chessboard trajectory. The x-axis is horizontal and
the t-axis  vertical. The sign of the contribution changes every two
corners in the trajectory. This is indicated in the figure by the
different line widths in the different segments.\label{twins1}}
 \end{figure}

 A lattice approximation to the Kernel
$K(b,a)$ for a particle to propagate from position $a$ at time
$t_a$ to position $b$ at time
$t_b$  is given by Feynman to be: 
\begin{equation}\label{nofr}
 K (b,a)=\sum_R N(R)(i\epsilon m)^R
\end{equation} where the sum is over all Chessboard paths and N(R) is the
number of paths with R corners. Here $m$ is the mass of the particle in
units where $\hbar$ is one. In terms of the paths themselves, the
expected distance between corners is $1/m$ \cite{JacSchulman84}. If we
distinguish between the two directions in space,
$K$ is  a 2x2 matrix which converges to the Dirac propagator in the
continuum limit\cite{Gersch81}.  The prescription given in (\ref{nofr})
can be modified somewhat for convenience. Gersch, who established the
relation between the Chessboard model and the one dimensional Ising
model, pointed out that the non-relativistic limit is more
direct if $i$ is replaced by $-i$ in (\ref{nofr}). Kull and
Treumann\cite{Kull99}  also noted that  paths fixed at both ends have
$(R-1)$ degrees of freedom, so the $R$ in (\ref{nofr}) may be replaced by
$(R-1)$ without interfering with the continuum limit.

 Equation (\ref{nofr}) is a formal analytic continuation (quantization)
of a classical partition function. The $i$ in the sum, which replaces a
real positive weight in the partition function, enforces the
quantization. It also partitions the sum into 4 components, each of which
is real, i.e.: 
\begin{eqnarray}\label{foursum}\nonumber
 K(b,a)&=&(\sum_{R=0,4,\ldots} N(R)(\epsilon m)^R
-\sum_{R=2,6,\ldots} N(R)(\epsilon m)^R)\\\nonumber 
     & +& i(\sum_{R=1,5\ldots} N(R)(\epsilon m)^R
-\sum_{R=3,7,\ldots} N(R)(\epsilon m)^R)\\
	&=&\Phi_R +i\,  \Phi_I.
\end{eqnarray} 
Each of the above sums is, by itself, a partition
function for a class of random walks in which the term $(\epsilon m)^R$ is just
a Boltzman weight. The interference of alternative paths is a result of the
two subtractions in (\ref{foursum}).  If we replace the minus signs
in (\ref{foursum}) by plus signs, the resulting propagator is related to the
Telegraph equation, which in turn becomes the diffusion equation in the
appropriate `non-relativistic' limit\cite{gord92chess}, the remaining $i$ then
being superfluous. The underlying stochastic model for this case has been
studied by Kac\cite{Kac74} and its relation to the Dirac equation through
analytic continuation has been discussed by Gaveau et.\/al.\cite{GJacobson84}
and Jacobson and Schulman\cite{JacSchulman84}. With the original minus signs
inplace, the
$i$ which appears in (\ref{foursum}) just expresses $K$ as a
particularly convenient linear combination of the real amplitudes
$\Phi_{R/I} $, however the actual interference characteristic of
quantization is apparent in the oscillatory nature of the
$\Phi $ themselves.

 Since it is the occurrence of the minus signs in the propagator which is
essential for interference we look for a physical basis for the
subtractions. 
 Regarding Fig.1 we can encode the counting and subtractions involved in
(\ref{foursum}) by colouring the trajectories with  two colours, say blue
(thick lines in figure) and red(thin in figure). If the trajectories start out
blue, they change to red at the second corner,  blue at the fourth and so on.
The sign of the contribution of a trajectory is then determined by its colour at
the end point, + for blue, --  for red. Red contributions behave like
antiparticles in that they reduce the contribution of the particles, providing
interference effects. The ensemble of such coloured paths between
$a$ and $b$ provides the appropriate contribution to a quantum
propagator, but is not explicitly traversed as a single path. What we
would like to do is to sew together the Chessboard paths in such a way
that they may be traversed by a single path which also provides the
alternating colours of the trajectories through the direction in time of
the traversal.  To this end, we note  from
Fig. 2  that  each Chessboard path has an orthogonal twin. 

\begin{figure}
\includegraphics[scale = .42]{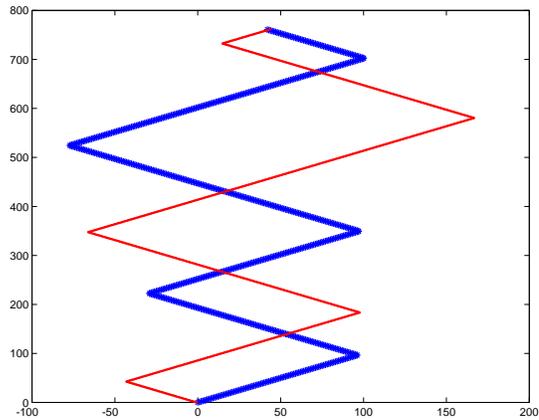}
\caption{The Chessboard Trajectory of Fig. 1  and its  Orthogonal Twin.
This pair can be viewed as two osculating Chessboard paths which never
cross, or as a single entwined loop which crosses itself
frequently. The latter view explains the phase shift of $\pi$ for every
two corners in the Chessboard paths.
\label{twins2}}
 \end{figure}
The orthogonal twin starts from the origin moving in
the opposite direction with the opposite colour. It moves the same
distance as the second leg of its twin's path, reverses direction and
moves the same distance as its twin's first leg. Twins meet at every
second corner where they change both colour and direction.   For paths with an
odd number of corners, this is repeated until the twins meet at  $t=t_b$ (for
paths with an even number of corners see below). The orthogonal twin is also a
Chessboard path with colouring
$180^o$ out of phase with the original.  

Now consider the following `entwined' traversal of the two paths. Follow
the first twin to the first meeting, the second to the second meeting and
so on. This path is blue from the origin to the last meeting. From there reverse
the direction in
$t$ by proceeding down the remaining red sections. This brings you back to
the origin on an entirely red path. This choice of traversal gives a
meaning to the original Feynman colouring; the colouring corresponds to
the direction in time of an {\em entwined} path traversal. Blue
corresponds to  forward in $t$, red to backwards. Entwined
pairs also conserve charge if we associate opposite charges with reversed time
segments. 

Each chessboard trajectory in (\ref{foursum}) has a unique
orthogonal twin. Let
$P_R$ be an arbitrary
$n$-step $R$-cornered Chessboard path. Write $P_R=(\sigma_1,\sigma_2,
\ldots,\sigma_n)$ where $\sigma_k =\pm 1$ according to the direction of the
$k$-th step of the path. If we define a `leg' as a set of
contiguous steps all in the same direction and bounded by either  corners
or ends of a path (i.e. a domain in the Ising analogy), then if $R$ is odd, we
may write
$P_R=(l_1,l_2,
\ldots,l_{R+1})$  with the understanding that $l_1$ stands for the
first leg, $l_2$ stands for the second and so on. If $ R$ is even then the path
ends with the last link in the same direction as the first link. In order to
join the path to an orthogonal twin we need to add a final leg in the opposite
direction. To do this uniquely we add a final leg the same length as the
original last leg but in the opposite direction. Thus  if $R$ is even, we
extend the n-step path to
$P_R=(l_1,l_2,
\ldots,l_{R+1},-l_{R+1})$, where $-l_{R+1}$ is $l_{R+1}$ with the signs of all
the component  $\sigma$ changed.  We may then define the orthogonal twin to
$P_R$ as
\begin{equation}  
 P_R^{\dagger} = \left\{ \begin{array}{llll}
(-l_1,l_1)& R \; = 0\\
(l_2,l_1)& R \; = 1\\
(l_2,l_1, \ldots,l_{R},l_{R-1},-l_{R+1},l_{R+1} ) & R=2,4,\ldots\\
(l_2,l_1,l_4,l_3, \ldots,l_{R+1},l_R ) & R =3,5, \ldots\\          
 \end{array} \right. 
\end{equation}
  Because
$P_R^\dagger$ is a unique permutation of
$P_R$, the ensemble, ${\cal
E}_F$, of all extended n-step paths
$P_R$ from the origin is the same as the ensemble of all paths
$P_R^\dagger$ from the origin. Furthermore, this is the same as the
ensemble of paths of the form $(+1,\sigma_2, \sigma_3 \ldots)$ combined
with all orthogonal twins. Thus we may cover all paths in ${\cal E}_F$,
with the correct Chessboard colouring, just by traversing all entwined
pairs. This may be done through a single continuous (in the sense of the
lattice) path since all entwined loops intersect at the origin.
Furthermore, entwined pairs fixed at the origin and at time $t_b$ have the
same number of degrees of freedom as their individual component
Chessboard paths (i.e. $R-1$) and
 each pair may be given the statistical weight $(\epsilon m)^{R-1}$ which
correctly weights the component Chessboard trajectories. Thus the
following classical stochastic process gives rise to a properly weighted
chessboard ensemble of coloured paths. Start a random walk at the origin
and allow the walker to choose entwined paths according to the number of
free corners, either in the entwined path or one of the pairs. The walker
traverses the entwined path as above  so as to maintain both the
Chessboard
 and time-sense colouring. The walker ends up at the origin at the end of
the traversal and repeats the process. The space-time lattice records the
net number of traversals in the $+t$ direction as the walker passes by
registering a plus one for a positive traversal and a minus one for a
negative traversal, thus accumulating positive and negative integers. 
The traversal weighting ensures that the constituent Chessboard paths
have the correct expected weight, and the ergodic nature of the walks
insures that, with enough loops, you can get as  close as you like to a
uniform coverage of the ensemble.   

If we allow a walker to cycle through the entwined paths according to the
above prescription, we can immediately write down the expected net `charge'
accumulated on the lattice. Referring to the kernel in (\ref{nofr}), we
can define the four components of the 2x2 matrix as $K_{\sigma_n
\sigma_1}$ where the subscripts refer to the end and beginning directions
respectively. If the walker, starting in the $+x$ direction, loops over
$N$ entwined pairs and $(x,t)$ is a lattice point within the light cone
with
$t<t_b$ then the contribution to the $+x$-component of the net charge is
proportional to
$\rho_+ = N\left(K_{++}(x,t)-K_{+-}(x,t)\right)$. This is because an entwined
loop corresponds to two forward Chessboard paths, one originating from the
origin with a positive, blue first leg ($K_{++}$ contribution)  and the second
from a negative red first leg($-K_{+-}$ contribution). Similarly the
$-x$-component at
$(x,t)$ is proportional to
$\rho_- =N\left(K_{-+}(x,t)-K_{--}(x,t)\right)$. The $\rho$ may be
interpreted as particle densities which may be either positive or
negative depending on the predominance of entwined trajectories in plus or minus
$t$ directions.
$\rho_+$ is positive in $(+x,+t)$-rich areas and $\rho_-$ is
positive in  $(-x,+t)$-rich areas.  Note that
$\rho_+ -\rho_-$ is proportional to the sum of the real and imaginary part of
the Feynman propagator $ \Phi_R +\,  \Phi_I$(Gersch convention for the sign of
$i$).
Unlike the predecessors of this
model\cite{gord92,McKeonOrd92,gord93lett}, which did not use bound pairs
of trajectories, this new model is relatively easy to simulate
on a lattice.

%
\begin{figure}
\includegraphics[scale = .82 ]{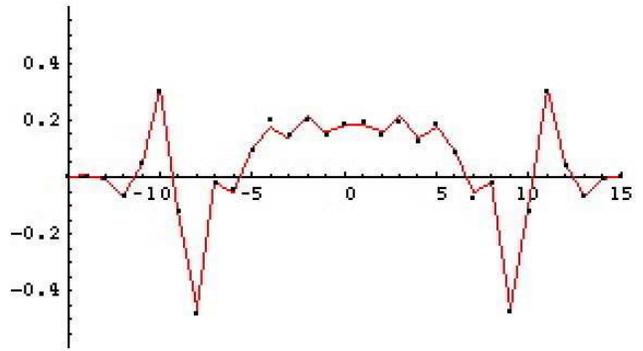}
\caption{ The sum of propagator components, $\Phi_R +\,  \Phi_I$ along the
$x$-axis, from the Chessboard model (curve) and the single path simulation
(points) at $t=15$ steps from the origin.\label{numprop}}
 \end{figure}
 Fig.(3) shows an example of such a simulation, where the sum of the
real and imaginary  parts of the propagator, $\Phi_R +\,  \Phi_I$, at fixed
$t$, are plotted versus $x$. The expected results from the Chessboard model
are plotted (continuous curve) at the same lattice resolution  as the results of
a simulation with a single path which loops over the lattice
$10^8$ times. In the figure, $t$ is 15 steps from the origin, with an average of
two steps between corners or a probability of $1/2$ for a direction change at
each step. At smaller values of
$t$, the simulation  and the Chessboard model are indistinguishable on the
scale of the figure, at larger values of $t$ the single path gives sparser
coverage of the chessboard ensemble and the scatter increases. The
individual real and imaginary parts of the propagator may be calculated
using the symmetry of the solutions, or by recording the $\rho_{\pm}$ in two
components to separate contributions from the original chessboard
path and its orthogonal twin. 

  Although we do not know how much of the above can survive inclusion of an
external field and/or  extension to three space dimensions, we do think the
 result reveals several qualitatively appealing features of the simplest case of
a free particle in one dimension.  First, the Feynman propagator  has an
independent existence as an expected net charge over an ensemble of entwined
paths which can be joined into a single trajectory. In this  context, the
propagator has an underlying classical stochastic  model which is in effect
{\em self-quantizing} and produces  real densities in place of amplitudes. 

A second feature is that the above model provides a bridge between
two distinct views of quantum mechanics in this case. Regarding
Fig. 2, we may view the two trajectories in three ways. We can consider 
them as two separate chessboard trajectories, coloured according to
Feynman's corner rule.  An ensemble of such trajectories  builds a
quantum propagator as a sum-over-histories. This is the conventional view.
A second picture is to note that an entwined pair forms a chain
of creation/annihilation events. An ensemble of these would provide a
vacuum of virtual particles upon which an excitation could presumably
propagate. This is close to a field theory perspective.

The third picture, which is suggested by the new formulation, is the
continuous loop in space-time, coloured according to direction of motion
in time. In this
picture, the phase of the wave function, `zitterbewegung', and the
presence of virtual particles are all manifestations of a single path which
forms  entwined space-time loops. In many respects, this picture is an
implementation of the original  Wheeler-Feynman
one-electron-universe\cite{Schweber94}, scaled down to provide a single-path
electron. Here the multiple tracks in space-time create a `Dirac sea' rather
than the multitude of electrons in the universe.

Finally, the entwined path formulation allows an analog  of
wavefunction collapse to be associated with the system. Suppose that we impose a
minimal requirement that a `measurement' at time $t=t_m$ must fix the
wavefunction at all times $t<t_m$. This requirement eventually(in a local
`time' parameter of the particle which measures distance along the full
space-time trajectory) forces the point particle which draws the propagator to
stay in the region
$t>t_m$, thus making it  redraw the `future'  propagator in a manner that is
consistent with an  `initial' condition at $t=t_m$. This change in the
wavefunction at $t_m$ need not be unitary and may provide the analog of
collapse.  An interesting next step would be to see if  a traversal and
measurement  scheme could be found that would initiate collapse  in a manner
consistent with the Born postulate.

\begin{acknowledgments} This work was partly funded by NSERC (GNO).
The authors are grateful for helpful discussions and computational
expertise  from John Dorband and Scott Antonille at NASA-GSFC. 
\end{acknowledgments}


\end{document}